\documentclass[pra, aps, twocolumn, floatfix, showpacs]{revtex4-1}
\usepackage{graphicx, amsmath, amssymb, times}

\usepackage{times}
\newcommand{\bk}{\mathbf{k}}
\newcommand{\br}{\mathbf{r}}

\begin{document}
\title{Rotating a Rashba-coupled Fermi gas in two dimensions}
\author{E. Doko$^{1}$, A. L. Suba{\c s}{\i}$^{2}$, and M. Iskin$^{1}$}

\affiliation{ 
$^{1}$Department of Physics, Ko{\c c} University, Rumelifeneri Yolu,
34450 Sar{\i}yer, Istanbul, Turkey. \\
$^{2}$Department of Physics, Faculty of Science and Letters, Istanbul
Technical University, 34469 Maslak, Istanbul, Turkey. 
}
\date{\today}
\begin{abstract}
We analyze the interplay of adiabatic rotation and Rashba spin-orbit coupling on 
the BCS-BEC evolution of a harmonically-trapped Fermi gas in two dimensions
under the assumption that vortices are not excited.
First, by taking the trapping potential into account via both the semi-classical
and exact quantum-mechanical approaches, we firmly establish the parameter 
regime where the non-interacting gas forms a ring-shaped annulus. 
Then, by taking the interactions into account via the BCS mean-field approximation, 
we study the pair-breaking mechanism that is induced by rotation, \textit{i.e.}, 
the Coriolis effects. In particular, we show that the interplay allows for the possibility 
of creating either an isolated annulus of rigidly-rotating normal particles that is 
disconnected from the central core of non-rotating superfluid pairs or an 
intermediate mediator phase where the superfluid pairs and normal particles 
coexist as a partially-rotating gapless superfluid. 
\end{abstract}
\maketitle

\section{Introduction}
\label{sec:introduction}

The quantum behavior of superfluids is most evident when they are placed 
in a rotating container. While a slow rotation may lead to the appearance of 
quantized vortices or quenching of the moment of inertia~\cite{Abo2001, Zwerlein2005}, 
rapidly-rotating systems may exhibit integer and fractional quantum-Hall 
physics~\cite{Ho2000, Cooper2001, Fischer2003, Baranov2005}. 
On the other hand, the Rashba spin-orbit coupling (SOC) involves an 
intrinsic angular momentum, caused by coupling the particle's spin to its
momentum, and it has become one of the key themes in modern condensed-matter 
and atomic physics, playing a central role for systems such as topological insulators 
and superconductors~\cite{Hasan2010, Qi2011}, quantum spin-Hall 
systems~\cite{Sinova2004}, and spintronics applications~\cite{Zutic2004}. 

Hence, the interplay between rotation and SOC is expected to pave the way 
for novel quantum phenomena, and recent realizations of SOC in atomic 
gases have already set the stage 
for such a task in a highly-controllable setting~\cite{Lin2011, Zhang2012, Wang2012, Cheuk2012, Qu2013, 
Goldman2014, Olson2014, Jimenez-Garcia2015,Huang2015}. 
While only the NIST-type SOC has so far been achieved, there also exist
various proposals on how to create a purely Rashba SOC~\cite{Ruseckas2005, 
Campbell2011, Xu2012}, the possibility of which has stimulated numerous 
theoretical studies on Rashba-Fermi gases in three~\cite{Vyasanakere2011, 
Jiang2001, Yu2011, Gong2011, Iskin2011, Yi2011, Zhou2012, Liao2012, Han2012} 
as well as two~\cite{He2012, Gong2012, Yang2012, Takei2012, 
Ambrosetti2014, Zhang2013, Iskin2013, Cao2014} dimensions. These works 
have revealed a plethora of intriguing phenomena, including topological 
superfluids, Majorana modes, spin textures, skyrmions, etc., which 
may soon be observed given the recent advances in producing a two-dimensional 
Fermi gas~\cite{Huang2015,Martiyanov2010, Dyke2011, Frohlich2011, Feld2011, Sommer2012, Ries2015}.

In this paper, we study the cooperation of adiabatic rotation and Rashba SOC 
on the ground-state phases of a trapped Fermi gas assuming that vortices are 
not excited. Adiabaticity requires that the rotation is introduced slowly to the system.
 In particular the rate of change of rotation frequency should be much smaller 
than the quasiparticle excitation frequency for vortex creation.
 In the absence of a SOC, effects of rotation on three-dimensional 
Fermi gases have previously been studied under this assumption at unitarity 
using the quantum Monte Carlo equations of state together with the 
local-density approximation (LDA)~\cite{Bausmerth2008a, Bausmerth2008}, 
and throughout the BCS-BEC evolution using the BCS mean-field 
approximation together both with 
LDA~\cite{Urban2008} and the Bogoliubov-de Gennes approach~\cite{Iskin2009}. 
These works have shown that, by breaking some of the superfluid (SF) pairs via
the Coriolis effects, rotation gives rise to a phase separation between a 
non-rotating SF core at the center and a rigidly-rotating normal (N) particles 
at the outer edge~\cite{Bausmerth2008a, Bausmerth2008}, with a partially-rotating 
gapless SF (gSF) region in between where the SF pairs and N particles coexist 
together~\cite{Urban2008, Iskin2009}. Since the effects of the Coriolis force on 
a neutral particle in the rotating frame are similar to those of the 
Lorentz force on a charged particle in a magnetic field, current advances in 
simulating artificial fields with ultracold atoms opens alternative ways of 
effectively realizing a rotating Fermi gas with SOC.
In addition, more recent works have confirmed that pair-breaking scenario is 
energetically more favored against vortex formation in a sizeable parameter 
regime~\cite{Warringa2011, Warringa2012}, and experimental schemes for 
realizing a rotating spin-orbit coupled system are described in ~\cite{Radic2011}.

\begin{figure}[htpb]
	\centering
    \includegraphics[scale=0.82]{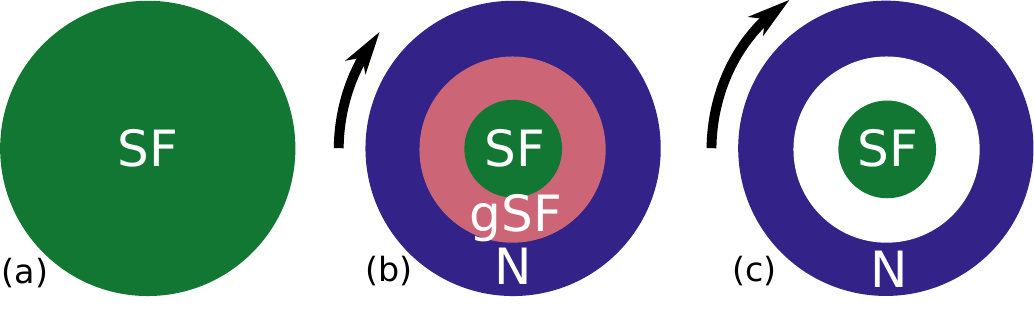} 
    \caption{ (Color online) Cartoon pictures showing that
	  (a) the whole gas is a SF in the absence of rotation,
	  (b) an intermediate gSF phase (where the SF and N particles coexist) 
	  emerges due to pair-breaking induced by sufficiently high rotations, and
	  (c) N particles is disconnected from the central SF core due to SOC 
	  at sufficiently high rotations.
    }
	\label{fig:phases}
\end{figure}

In the presence of a Rashba SOC, we first take the harmonic confinement into 
account via both the semi-classical LDA and exact approaches, 
and find the parameter regime where the non-interacting 
gas forms a ring-shaped annulus. Neither rotation nor SOC alone can deplete 
the central density to zero no matter how fast the rotation or large SOC is, 
and the formation of such an intriguing annulus requires them both simultaneously.
Then, we take the interactions into account via the BCS mean-field, and analyze 
the pair-breaking mechanism in the entire BCS-BEC evolution. In particular, 
we show that the cooperation of rotation and SOC allows for the possibility of 
creating either an isolated annulus of N particles that is disconnected from the 
central SF core or separated from it by an intermediate gSF phase as sketched in 
Fig.~\ref{fig:phases}. We also construct phase diagrams showing the very first 
emergence of an N phase and the complete destruction of the central SF core.

\section{Theoretical Formalism}
\label{sec:TF} 

To obtain these results, we use the following Hamiltonian density in the rotating frame
\begin{align}
& H (\mathbf{r}) = \psi^{\dagger}(\br) 
\left[ 
\frac{\mathbf{p}^{2}}{2M}+V(r)-\mu -\Omega L_z +\alpha \mathbf{p}\cdot\vec{\mathbf{\sigma}} 
\right] \psi(\br) \nonumber \\
&+ \Delta(\boldsymbol{r}) \psi_{\uparrow}^{\dagger}(\br)\psi_{\downarrow}^{\dagger}(\br)
+ \Delta^*(\boldsymbol{r}) \psi_{\downarrow}(\br)\psi_{\uparrow}(\br) 
+ \frac{|\Delta(\boldsymbol{r})|^2}{g},
\end{align}
where $\psi^\dag=(\psi_{\uparrow}^\dag,\psi_{\downarrow}^\dag)$ denotes 
the field operators, $\mathbf{p}=-i\mathbf{\nabla}$ is the linear-momentum operator 
with $\hbar = 1$, $V(r)=M\omega^{2}r^{2}/2$ is the trapping potential,
$\mu$ is the chemical potential, $0 \le \Omega < \omega$ is the rotation frequency, 
$L_z$ is the $z$-projection of the angular-momentum operator 
$\mathbf{L} = \mathbf{r} \times \mathbf{p}$,
$\alpha \ge 0$ is the strength of the Rashba coupling,
$\vec{\mathbf{\sigma}}=(\sigma_x, \sigma_y)$ is a vector of Pauli spin matrices, and
$\Delta(\boldsymbol{r})=g\langle\psi_{\uparrow}(\br)\psi_{\downarrow}(\br)\rangle$ is 
the mean-field SF order parameter with $g \ge 0$ being the strength of attractive 
interactions and $\langle \cdots \rangle$ the thermal average. 
Within the semi-classical LDA, the local Hamiltonian can be written as 
$
H_\mathrm{loc} = (1/2) \sum_{\bk}\psi_{\bk}^{\dagger} H_{\bk} \psi_{\bk}+C,
$
where the matrix
\begin{equation}
H_{\bk}\!\!=\!\!\begin{pmatrix}
\!\xi_{\bk}-\Omega L_{\bk} & S_{\bk} & 0 & \Delta\\
S_{\bk}^{*} & \!\!\xi_{\bk}-\Omega L_{\bk} & -\Delta & 0\\
0 & -\Delta^{*} & \!\!\!-\xi_{\bk}-\Omega L_{\bk} & S_{\bk}^{*}\\
\Delta^{*} & 0 & S_{\bk} &\!\!\! -\xi_{\bk}-\Omega L_{\bk}
\end{pmatrix}
\label{eqn:ham}
\end{equation}
governs the dynamics. The index $\mathbf{r}$ is dropped here and 
throughout for notational simplicity. In momentum space,
$
\psi_{\bk}^{\dagger}=(a_{\bk \uparrow}^{\dagger},a_{\bk \downarrow}^{\dagger},
a_{\mathbf{-k} \uparrow},a_{\mathbf{-k} \downarrow})
$
and $a_{\bk,\sigma}$ ($a_{\bk,\sigma}^{\dag}$) annihilates (creates) a $\sigma$ 
fermion with momentum $\bk = (k_x, k_y)$. 
The free-particle dispersion relative to the local chemical potential 
$\mu_\mathbf{r} = \mu - V(r)$ is 
$\xi_{\bk}=\epsilon_{\bk} - \mu_\mathbf{r}$
with $\epsilon_{\bk}=k^{2}/(2M)$.
The rotation enters via $\Omega L_{\bk} = \mathbf{v}(\br)\cdot\bk$ with 
the velocity $\mathbf{v}(\br)=\Omega\mathbf{\hat{z}}\times\br$,
$S_{\bk}=\alpha (k_{x}-ik_{y})$ is the SOC term, 
$\Delta = g \sum_{\bk}\langle a_{\bk \uparrow}a_{-\bk \downarrow}\rangle$
denotes the SF order parameter, and 
$
C=\sum_{\bk}(\xi_{\bk}+\Omega L_{\bk})+|\Delta|^{2}/g
$ 
is a constant.

We diagonalize Eq.~(\ref{eqn:ham}) and express
$
H_\mathrm{loc} = C + \sum_{\bk s} ( E_{\bk s}\gamma_{\bk s}^{\dag}\gamma_{\bk s}-E_{\bk s}/2 ),
$
where $\gamma_{\bk s}^{\dag}$ ($\gamma_{\bk s}$) creates (annihilates) a quasiparticle 
with helicity $s=\pm$ and energy
$
E_{\mathbf{k} s}=\sqrt{(\xi_{\bk}+s\alpha k)^{2}+|\Delta|^{2}}-\Omega L_{\bk}.
$
Thus, the thermodynamic potential 
$
G = - (1/\beta) \mathrm{Tr}(\ln e^{-\beta H_\mathrm{loc}}),
$ 
where $\beta=1/(k_B T)$ with $k_B$ the Boltzmann constant and $T$ the temperature, 
can be written as 
$
G = C + \sum_{\bk s}[E_{\mathbf{k} s}f(E_{\bk s}) - E_{\bk s}/2]
$
with the Fermi function $f(x)=1/(e^{\beta x}+1)$. 
Setting $\partial G/\partial|\Delta| = 0$ and using 
$n(\mathbf{r}) = - (1/A) \partial G/\partial\mu_\mathbf{r}$ for the local particle 
density in area $A$, we finally obtain the local self-consistency equations
\begin{align}
\label{eq:Gap}
\frac{1}{g} &=\frac{1}{4}\sum_{\bk s}\frac{1}{E_{\bk s}+\Omega L_{\bk}} \tanh(\beta E_{\bk s}/2), \\
\label{eq:Number} 
n(\br) &= \frac{1}{2 A} \sum_{\bk s}
\left[1-\frac{\xi_{\bk}+s\alpha k}{E_{\bk s}+\Omega L_{\bk}} \tanh(\beta E_{\bk s}/2)\right],
\end{align}
such that the total number of particles is given by $N=\int d\br n(\br)$. 
These equations are the generalizations of the usual BCS expressions to the 
case of rotation and SOC, and it is a standard practice to substitute the two-body 
binding energy in vacuum $E_{b} \ge 0$ for $g$ via the relation
$
1/g=\sum_{\textbf{k}}1/(2\epsilon_{\textbf{k}}+E_b)
$
in the cold-atom context.
While a non-zero (vanishing) $\Delta$ is a characteristic of SF (N) phase in general, 
we also use the mass-current density $\mathbf{J} = (J_x, J_y)$ flowing in the
azimuthal direction, where
\begin{align}
(J_x, J_y) &= \frac{1}{A} \sum_{\bk}
\left[ n_{\bk} \bk +  2M\alpha \left(  \Re \lbrace p_{\bk} \rbrace, \Im \lbrace p_{\bk} \rbrace \right) \right],
\end{align}
to further characterize the gSF phase. Here,
$
n_{\bk} =  \langle a^{\dag}_{\bk \uparrow}a_{\bk \uparrow} \rangle 
+ \langle a^{\dag}_{\bk \downarrow}a_{\bk \downarrow} \rangle
$
is the momentum distribution given by the summand of Eq.~(\ref{eq:Number}),
$\Re$ and $\Im$ denote the real and imaginary parts and
$
p_{\bk} = \langle a^{\dag}_{\bk \uparrow} a_{\bk \downarrow} \rangle.
$
Let us first set $\Delta = 0$ and $T = 0$, and analyze the non-interacting ground state.

\section{Non-interacting Problem}
\label{sec:NI} 

In the absence of both rotation and SOC, the non-interacting gas has the shape of 
a disc with its density $n(r)$ decreasing parabolically as a function of $r$, until to the
edge of the system given by the Thomas-Fermi radius $R_F=\sqrt{2E_F/(M\omega^2)}$, 
where $E_F = k_F^2/(2M) = \omega \sqrt{N}$ is the Fermi energy. 
In the presence of rotation only, while the gas retains its overall parabolic density, 
some of the the particles are expelled from the center of the trap due to the centrifugal 
effects, and the edge moves to $\widetilde{R}_F=R_F/(1-\Omega^2/\omega^2)^{1/4}$.
However, in the presence of SOC only, $n(r)$ tends to increase at the center and 
the gas shrinks due to the increased low-energy density of states. Therefore, 
$\Omega \ne 0$ and $\alpha \ne 0$ have counteracting effects on $n(r)$ in general. 
In addition, since $\Omega \ne 0$ causes an asymmetry in the local $\bk$ space 
and $\alpha \ne 0$ shifts its dispersion minima from $\mathbf{k} = \mathbf{0}$ to 
$\mathbf{k} \ne \mathbf{0}$, their interplay is expected to give rise to a much richer 
physics even in the non-interacting limit. For instance, setting $\Delta = 0$ in 
Eq.~(\ref{eq:Number}), and solving for $n(r) = 0$, we find an analytic expression 
for the edge(s) of the gas given by
\begin{equation}
R_{O, I}^0 = R_{F} \frac{\omega\Omega\alpha \pm \omega
\sqrt{\alpha^{2}\omega^{2}+2\mu(\omega^{2} -\Omega^{2})/M}}
{k_{F}(\omega^{2}-\Omega^{2})/M},
\label{eq:RE}
\end{equation}
where $R_{O}^0$ ($R_{I}^0$) is the radius of the outer (inner) edge. 
Note that while $R_O^0$ is positive and exists for all parameters as long as 
$\Omega < \omega$, a positive $R_I^0$ is possible only for the parameter regimes 
where $\alpha^{2}+2\mu/M < 0$. While the gas has 
the usual shape of a disc with $n(r) \ne 0$ for $0 \le r < R_O^0$ when 
$\alpha^{2} > -2\mu/M$, otherwise it has the shape of a ring with 
$n(r) \ne 0$ in the radial interval $R_I^0 < r < R_O^0$.

\begin{figure}
\includegraphics[trim=70 10 72 25, clip, scale=0.27]{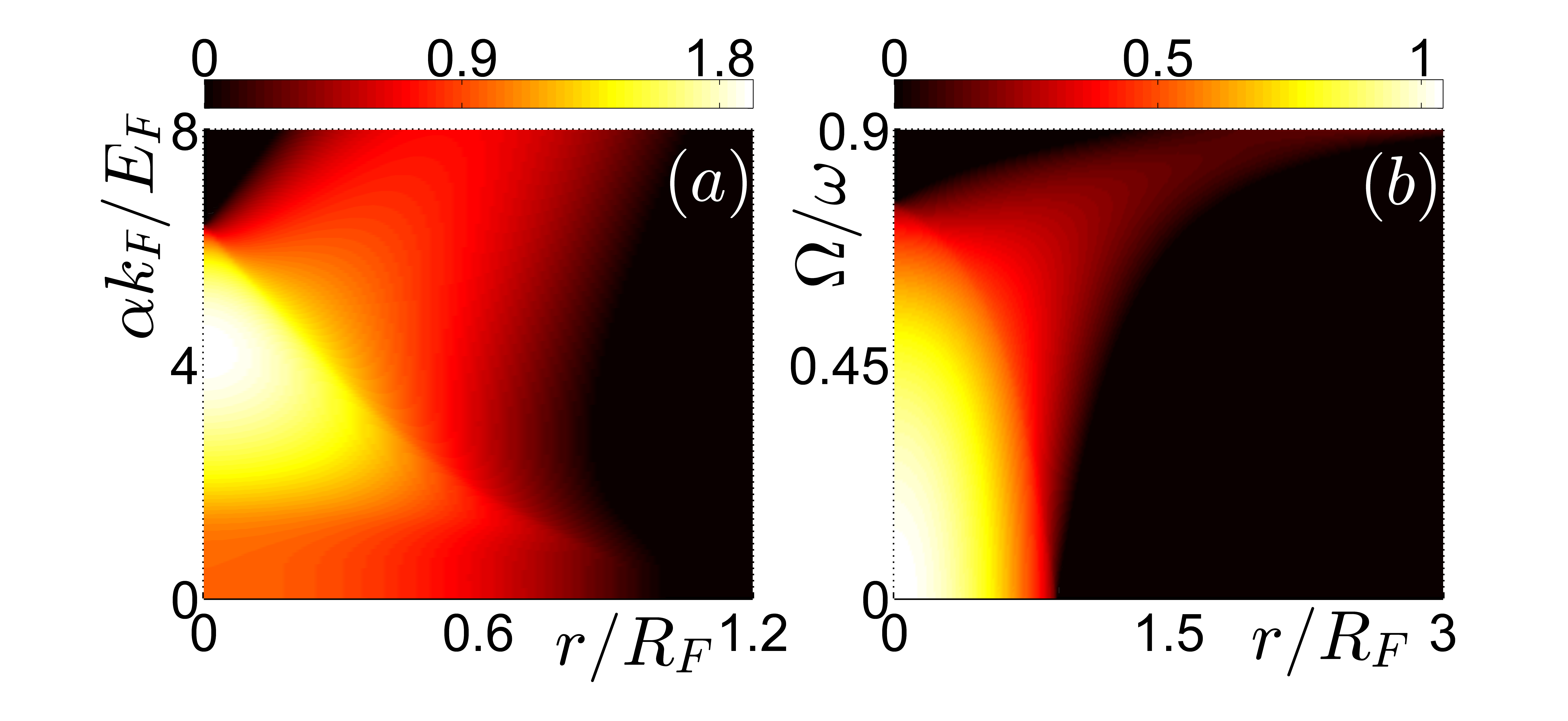} 
\caption{(Color online)
Non-interacting $n(r)$ profiles are shown as a function of
(a) $\alpha$ for $\Omega=0.15\omega$, and 
(b) $\Omega$ for $\alpha=1E_F/k_F$. 
The gas eventually forms a ring-shaped annulus in both figures.
\label{fig:freeN}}
\end{figure}

The formation of such an annulus can also be illustrated by solving $n(r)$ 
directly from Eq.~(\ref{eq:Number}), which we present in Fig.~\ref{fig:freeN}. 
As shown in Fig.~\ref{fig:freeN}(a), while increasing $\alpha$ initially 
increases $n(r)$ near the central region due to the increase in the low-energy 
density of states, the $\mathbf{k} \ne \mathbf{0}$ states gain further energy 
through their angular momentum in the rotating frame, and $n(r)$ decreases 
dramatically away from the center as $\alpha$ increases further. As the critical condition 
is approached, the latter effect gradually dominates causing central density to 
decrease as the gas continues to expand, and $n(0)$ eventually vanishes beyond
$\alpha=\sqrt{-2\mu/M}$. We also see a similar behavior in Fig.~\ref{fig:freeN}(b), 
where increasing $\Omega$ is shown to deplete $n(0)$ to zero once the critical condition 
is satisfied. This happens at faster $\Omega$ for smaller $\alpha$ and vice versa.
In contrast to increasing $\alpha$, however, increasing $\Omega$ 
not only decreases $n(0)$ but also expands the gas monotonically. It is important
to emphasize here that neither rotation nor SOC alone can deplete $n(0)$ 
to zero no matter how fast $\Omega$ or large $\alpha$ is, and the formation of an 
annulus requires them both simultaneously.

In the non-interacting limit, we benchmark our semi-classical results with those of 
exact quantum-mechanical treatment and find an excellent agreement between 
the two for all parameters including the fast $\Omega$ and/or large $\alpha$ limits. 
The details of this comparison are given in~Appendix~\ref{app:LDAvsQm}. 
Motivated by this success, next we apply the LDA formalism to the entire 
BCS-BEC evolution at $T = 0$.

\section{Interacting Problem}
\label{sec:IP} 

In the absence of both rotation and SOC, 
the superfluid persists with $\Delta \ne 0$ as long as $n(r) \ne 0$,
and therefore, the entire system is a disc-shaped gapped SF with its edge 
located at $R_{O} = R_{F}$
no matter how strong or weak $E_b$ is, 
which only happens in two dimensions. 
In the presence of an adiabatic rotation, 
since vortices are assumed not to be excited in the system and the 
gapped SF can not carry the angular momentum, some of the SF pairs must 
eventually break by the centrifugal effects, 
\textit{i.e.}, via the broken time-reversal symmetry, giving rise to gSF and N 
regions in the trap. In this paper, we distinguish gSF from SF by checking 
whether $J$ is non-zero or not, or equivalently whether $E_{\bk s}$ has negative 
regions in $\mathbf{k}$ space or not. In addition, while both gSF and N phases 
have $J \ne 0$, only the N region rotates rigidly as a whole with $J = M n(r) \Omega r$.

\begin{figure}
\includegraphics[scale=1.0]{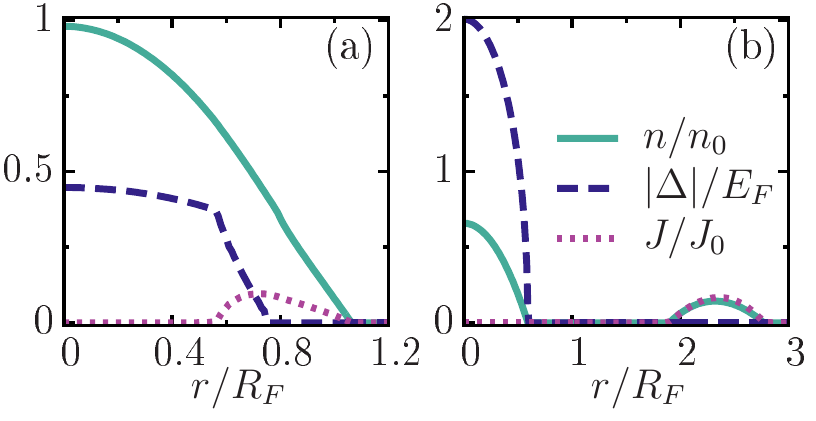} 
\caption{(Color online)
Typical profiles showing
(a) a partially-rotating gSF region, and
(b) a vacuum
seperating the non-rotating SF pairs from the rigidly-rotating N particles.
Here, $E_{b}=0.1E_{F}, \alpha=0.8 E_F/k_F$ and $\Omega=0.3\omega$ in (a),
and $E_{b}=1E_{F}, \alpha=7 E_F/k_F$ and $\Omega=0.5\omega$ in (b),
where $n_0 = k_F^2/(2\pi)$ and $J_0 = M n_0 \omega R_F$.
\label{fig:Ex}}
\end{figure}

When $\alpha = 0$, the robustness of the SF pairs against rotation depends 
on $E_b$ and $\Omega$ in such a way that there is no pair breaking 
when $\Omega$ is sufficiently slow for a given $E_b$. Therefore, $n(r)$, 
$\Delta$ and $\mu$ are not affected by rotation as long as $\Omega < \Omega_c$, 
and the entire system is again a disc-shaped SF with its edge located at 
$R_{O} = R_{F}$. We determine $\Omega_c$ by noting that when the first
broken pair emerges as N particles at the edge of the gas then its radius 
must coincide with the Thomas-Fermi radius of the non-rotating gas 
with the same $\mu$, \textit{i.e.}, 
$
R_{O}^{0}(\alpha \to 0, \Omega \to \Omega_{c}) = R_{F},
$
leading to $\Omega_{c}=\omega\sqrt{E_{b}/(2E_{F})}$. 
Since $\Omega < \omega$ has a physical upper bound in order not to 
loose the particles from the harmonic trap, the SF pairs are perfectly robust 
against rotation for $E_{b} > 2E_{F}$ in the $\alpha \to 0$ limit.
When $\Omega > \Omega_c$, the trap profile in general consists of three 
distinct regions: while the central (outer) core (wing) is solely occupied by 
paired (unpaired) SF (N) particles, the SF pairs and N particles coexist in the 
middle as a result of which the gSF emerges as an intermediate phase 
around the SF-N interface. We note that even though the SF core shrinks 
monotonically and gives way to N phase with increasing $\Omega$, it still 
survives in the $\Omega \to \omega$ limit.

The preceding description still holds when $\alpha \ne 0$ but small, and we illustrate
a typical trap profile in Fig.~\ref{fig:Ex}(a), where a small gSF region is clearly 
visible at the SF-N interface. Similar to the $\alpha = 0$ case, the critical
rotation frequency for the emergence of N particles can be determined from 
$
R_{O}^{0}(\alpha,\Omega_{c}) = R_{O}(\alpha, \Omega = 0).
$
Away from the small-$\alpha$ limit, however, we find a very intriguing situation 
provided that $\alpha^{2} < -2\mu/M$. For instance, a typical trap profile 
for this case is shown in Fig.~\ref{fig:Ex}(b), where the N particles form an 
isolated annulus that is completely disconnected from the central 
SF core without a gSF region in between. The $\Omega_c$ threshold 
for the emergence of an isolated N phase can again be determined from
$R_{O}^{0}(\alpha,\Omega_{c}) > R_{O}(\alpha, \Omega = 0)$. 

\begin{figure}
\includegraphics[scale=1.05]{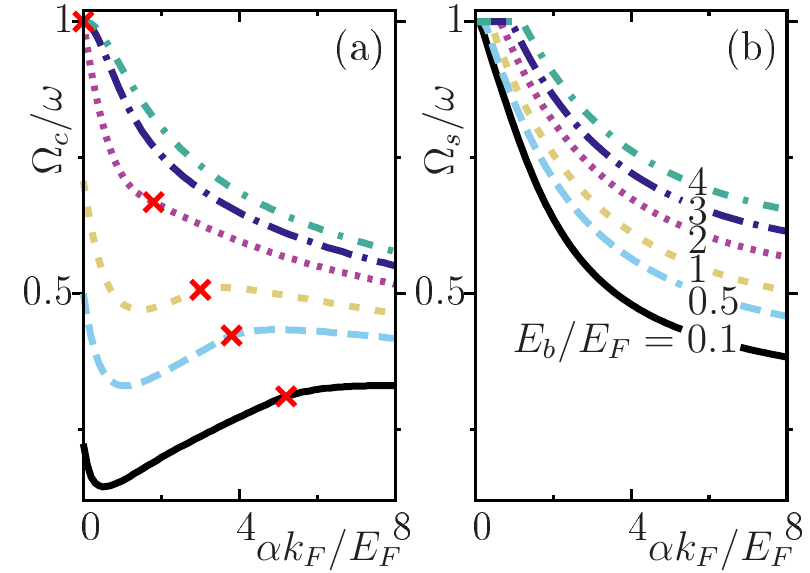} 
\caption{(Color online) Critical $\Omega$ curves are shown for
(a) the emergence of an N phase, and 
(b) the depletion of the SF core.
Beyond the (red) cross marks in (a), the N particles form an isolated annulus that is 
disconnected from the central SF core without a gSF region in between.
\label{fig:omc}}
\end{figure}

By repeating this analysis numerous times for a wide range of parameters, 
we construct the phase diagram of the system based on the very first 
emergence of N particles with increasing $\Omega$. The diagram shown 
in Fig.~\ref{fig:omc}(a) is one of our primary results in this paper and should 
be read as follows. For a given $\alpha$ shown on the horizontal axis, 
increasing $\Omega$ in the vertical direction breaks SF pairs beyond the 
intersection with the $\Omega_c$ curve. This diagram intuitively suggests 
that $\Omega_c$ increases with increasing $E_b$ for a given $\alpha$. 
More importantly, it also shows that, in contrast to the 
$\alpha \to 0$ limit where SF pairs are perfectly robust against $\Omega < \omega$ 
when $E_b > 2E_F$, finite $\alpha$ eventually allows $\Omega$ to 
create an N phase for any $E_b$ with $\Omega_{c} < \omega$. Furthermore, 
depending on whether the intersection with the $\Omega_c$ curve is to the 
left or to the right of the (red) cross marks, we know whether an intermediate gSF 
phase exists or not at the SF-N interface. In the former case, the gSF phase may 
eventually disappear with further increase in $\Omega$, so that the N phase ends 
up again being disconnected from the SF core (not shown in the phase diagram). 
Thus, the gSF phase always emerges for any $E_b < 2E_F$ 
in the $\alpha \to 0$ limit, and the N particles form an isolated annulus practically 
for any $E_b \gtrsim 2E_F$ as long as $\alpha \ne 0$. 
We note that increasing $E_b$ moves the cross marks to lower $\alpha$ 
because faster $\Omega_c$ leads to an annulus of N particles at smaller 
$\alpha$, as discussed for the non-interacting problem.

It is also possible to obtain an analytic expression for the small-$\alpha$ dependence 
of $\Omega_c$ on the left side of the cross marks, by again noting that the first 
broken pair emerges as N particles at the edge $r \to R_O$ of the gas. We evaluate 
the gapless condition $E_{\bk s} = 0$ with $\Delta \to 0$, after setting 
$
\mu \simeq E_{F} - E_{b}/2 - M\alpha^{2}
$ 
and $R_{O} \approx R_{F}$ at the lowest orders in $\alpha$, leading to
$
\Omega_{c} \approx \omega (\sqrt{2\alpha^{2}k_{F}^{2}/E_{F}^{2}+2E_{b}/E_{F}} - \alpha k_{F}/E_{F})/2.
$
This expression shows that $\Omega_c$ decreases with $\alpha$ at the lowest order,
and it is in excellent agreement with all of our numerical results.
Physically, since $\alpha \ne 0$ shifts the excitation minima to higher momenta,
the lowest-energy states become more susceptible to rotation, making
it easier for $\Omega$ to break the SF pairs at least in the small-$\alpha$ limit.
However, followed by a sudden decrease, Fig.~\ref{fig:omc}(a) also shows that
$\Omega_c$ curve develops a minimum as a function of $\alpha$ 
when $E_{b} \lesssim 2 E_F$. This is because of a competing mechanism 
where increasing $\alpha$ not only enhances $\Delta$ but also decreases 
$R_O$ by increasing $n(r)$ near the center, both of which make it more 
difficult for $\Omega$ to break the SF pairs away from the small-$\alpha$ limit. 
We also note that increasing $E_b$ moves the location of the minimum to 
larger $\alpha$ because it is only then the competing effects caused by SOC 
become comparable to the enhancement of $\Delta$ caused by stronger $E_b$.

\begin{figure}
\begin{center}
\includegraphics[scale=1.05]{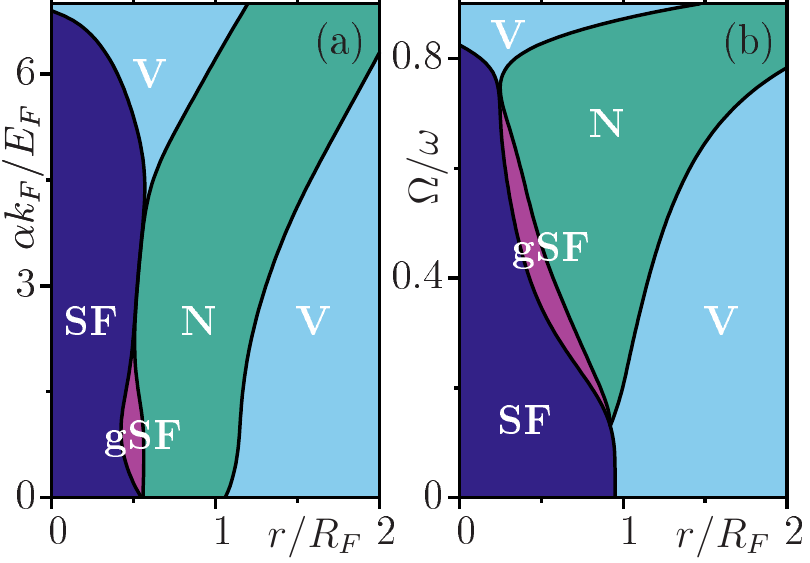} 
\end{center}
\caption{(Color online) 
Radial phase profiles are shown as a function of 
(a) $\alpha$ for $E_{b}=0.1E_{F}$ and $\Omega = 0.4\omega$, and 
(b) $\Omega$ for $E_{b}=0.1E_{F}$ and $\alpha=0.4 E_F/k_F$.
The cooperation of $\Omega$ and $\alpha$ not only destroys the central 
SF core but also boosts the gSF region in both figures.
\label{fig:gSF}
}
\end{figure}

Next to Fig.~\ref{fig:omc}(a), we present another phase diagram showing 
the destruction of the central SF core under more extreme parameter regimes. 
For instance, typical phase profiles for this case are illustrated in Fig.~\ref{fig:gSF}, 
where the interplay of fast $\Omega$ and/or large $\alpha$ eventually dominates 
the effects of $E_b \ne 0$ and depletes $n(0)$ to zero, recovering the 
non-interacting problem discussed above.
The diagram shown in Fig.~\ref{fig:omc}(b) is also one of our primary results 
in this paper, where $\Omega_s$ threshold for the complete destruction 
of the SF core is determined by first setting $\Delta \to 0$ as $r \to 0$ in 
Eq.~(\ref{eq:Gap}) to get $\mu$, and then plugging these values to 
Eq.~(\ref{eq:Number}). This diagram intuitively suggests that $\Omega_s$ 
increases with increasing $E_b$ for a given $\alpha$.
More importantly, it also shows that, in contrast to the $\alpha \to 0$ limit
where the central SF pairs are robust against $\Omega < \omega$ for any
$E_b$ since the centrifugal effects strictly vanish at $r = 0$, increasing 
$\alpha$ eventually allows $\Omega$ to destroy all of the SF pairs.

\section{Conclusions}
\label{sec:Conc} 

To summarize, here we studied the cooperation of adiabatic rotation and Rashba 
SOC on the ground-state phases of a trapped Fermi gas in two dimensions, 
assuming that vortices are not excited. First, by taking the harmonic confinement 
into account via both the LDA and exact approaches, we found the parameter 
regime where the non-interacting gas forms a ring-shaped annulus. 
It is important to emphasize that neither rotation nor SOC alone can deplete 
the central density to zero no matter how fast the rotation or large SOC is, 
and the formation of such an intriguing annulus requires them both 
simultaneously. Then, by taking the interactions into account via the BCS 
mean-field, we analyzed the rotation-induced pair-breaking mechanism 
in the entire BCS-BEC evolution. In particular, we showed that 
the cooperation of rotation and SOC allows for the possibility of creating 
either an isolated annulus of rigidly-rotating N particles that is disconnected from 
the central core of non-rotating SF pairs or an intermediate gapless SF phase
which is charecterized by the coexistence of SF pairs and N particles.
We also constructed phase diagrams showing the very first emergence of an 
N phase and the complete destruction of the central SF core. We  hope that,
given the ongoing push towards simulating Rashba-coupled Fermi gases 
by many groups worldwide, our compelling results may soon be realized once 
the technical experimental difficulties are cleared out of the way.

\section{Acknowledgments}

This work is supported by T\"{U}B$\dot{\mathrm{I}}$TAK Grant No. 1001-114F232 and
E. D. is supported by the T\"{U}B$\dot{\mathrm{I}}$TAK-2215 Ph.D. Fellowship.

\appendix
\section{Exact numerical solution for Sec.~\ref{sec:NI}}
\label{app:LDAvsQm}

In the rotating frame, the non-interacting Hamiltonian for a harmonically-trapped 
Fermi gas with Rashba SOC can be written as
\begin{equation}
H_{0}=\int d\br\, \psi^{\dagger}(\br) \left[ 
	H_\mathrm{HO}
-\Omega L_z -\mu+ 
\alpha \mathbf{p}\cdot\vec{\mathbf{\sigma}} 
\right] \psi(\br)\, ,
\end{equation}
where 
$
H_\mathrm{HO}=\mathbf{p}^{2}/(2M) + M \omega^2 r^2/2
$
is the usual two-dimensional harmonic-oscillator Hamiltonian. We make use of the 
rotational symmetry of the system, and expand the field operators in terms of the 
angular-momentum basis of the two-dimensional harmonic oscillator as
\begin{equation}
  \psi_\sigma (\br) = \sum_{n,l} 
    g_{n,l}(\br)
    c_{n,l,\sigma}\, ,
  \label{eq:psi_2dho}
\end{equation}
where $c_{n,l,\sigma}$ annihilates a spin $\sigma$ particle in the $|n,l \rangle$ 
state that is given by the normalized real-space wave function
\begin{equation}
  g_{n,l}(\br) = i^{2n_{_<}} \sqrt{\frac{n_{_<}!}{\pi a_0^2 n_{_>}!}} e^{il\theta} 
  \tilde{r}^{|l|} e^{-\tilde{r}^2/2}L^{|l|}_{n_{_<}}\left( \tilde{r}^2 \right)\; ,
  \label{eq:psinl}
\end{equation}
where $\tilde{r}=r/a_0$ with the harmonic-oscillator length $a_0 = \sqrt{1/(M \omega)}$ 
scale, the energy and angular-momentum quantum numbers $n \ge |l| \ge 0$ are 
integers, $n_{_<}/n_{_>}$ is the lesser/greater of $(n \pm l)/2$, and 
$
 L^{|m|}_k(x) = (x^{-|m|} e^x/k!) d^k ( e^{-x} x^{k+|m|} ) / dx^k.
$
is the $k^{\mathrm{th}}$-degree associated Laguerre polynomial.
The non-interacting Hamiltonian can be written in this basis as
\begin{eqnarray}
H_0&=&\sum_{n,l,\sigma}\left[ \omega (n+1)-\Omega l -\mu \right] c_{n,l,\sigma}^\dag c_{n,l,\sigma} 
\nonumber \\
&&+\frac{\alpha}{2a_0} \sum_{n,l} 
\left(i\sqrt{\frac{n+l}{2}+1}c_{n+1,l+1,\downarrow}^\dag c_{n,l,\uparrow} \right. 
\nonumber \\
&&\left.-i\sqrt{\frac{n-l}{2}}c_{n-1,l+1,\downarrow}^\dag c_{n,l,\uparrow}+ \textrm{H.c.}\right),
\end{eqnarray}
where H.c. is the Hermitian conjugate. Even though the Rashba coupling 
$\alpha\mathbf{p}\cdot\vec{\mathbf{\sigma}}$ does not preserve the individual spin 
or real-space rotational symmetry, the full rotational symmetry is still manifest. 
Since $H_0$ commutes with the total angular-momentum operator 
$
J_z=L_z+S_z= \sum_{n,l,\sigma} \left( l + \frac{\sigma_z}{2} \right) c_{n,l,\sigma}^\dag c_{n,l,\sigma},
$
where $\sigma_z=\pm 1$ for $\sigma=(\uparrow,\downarrow)$, respectively, 
they can be simultaneously diagonalized.

Using the conservation of $J_z$, the Hamiltonian can be expressed in a block-diagonal form,
$
H_0=\sum_{l} \Psi^\dag_l H_l \Psi_l,
$
where $H_l$ is a tri-diagonal matrix in each block with the following ordering of 
the basis states.
For $l \geq 0$ corresponding to the $J_z$ subspace ($l\uparrow, l+1\downarrow$)
with eigenvalue $j=l+1/2>0$, $H_l$ is given by
\begin{equation}
 \begin{pmatrix}
    \epsilon_{l,l}    & a\sqrt{l+1}     &  &  & \\
    a^*\sqrt{l+1} & \epsilon_{l+1,l+1}  &  a\sqrt{1}              &  &    \\ 
	          &  a^*\sqrt{1}    & \epsilon_{l+2,l} &  a\sqrt{l+2}  &    \\ 
	          &                 & a^*\sqrt{l+2} & \epsilon_{l+3,l+1} &  a\sqrt{2}  &   \\ 
	          &                 & & a^*\sqrt{2} & \epsilon_{l+4,l} &  \ddots \\    
	          &                 & & & \ddots & \ddots
  \end{pmatrix},
\end{equation}
where the basis vectors are ordered as
$
\Psi^\dag_l = (c^\dag_{n=l,l,\uparrow}, c^\dag_{l+1,l+1,\downarrow}, c^\dag_{l+2,l,\uparrow}, c^\dag_{l+3,l+1,\downarrow}, \dots ),
$
Here, $\epsilon_{n,l}=\omega(n+1)-\Omega l-\mu$ and $a=i\alpha/2a_0$. 
Similarly for $l<0$ corresponding to the ($-|l|+1\downarrow, -|l|\uparrow$) subspace
with eigenvalue $j=l+1/2<0$, $H_l$ is given by
\begin{equation}
  \begin{pmatrix}
    \epsilon_{-l-1,l+1} & a\sqrt{-l}    &  &  & \\
    a^*\sqrt{-l}    & \epsilon_{-l,l} &  a\sqrt{1}              &  &    \\ 
	            & a^*\sqrt{1}   & \epsilon_{-l+1,l+1} &  a\sqrt{-l+1}  &    \\ 
	            &               & a^*\sqrt{-l+1}  & \epsilon_{-l+2,l} & a\sqrt{2}  &   \\ 
	            &               &                 & a^*\sqrt{2}     & \epsilon_{-l+3,l+1}  & \ddots \\    
	            &               &                 &                 & \ddots & \ddots
\end{pmatrix},
\end{equation}
where the basis vectors are ordered as
$
\Psi^\dag_l = (c_{n=-l-1,l+1,\downarrow}, c_{-l,l,\uparrow}, c_{-l+1,l+1,\downarrow}, c_{-l+2,l,\uparrow}, \dots ).
$
This Hamiltonian can be diagonalized via the unitary transformation 
$
c_{n,l,\sigma} =\sum_{m}u^{(j)}_{n\sigma,m} C_{j,m},
$
leading to
$
H_0=\sum_{j,m} E^{(j)}_{m} C_{j,m}^\dag C_{j,m},
$
where $j=l+\sigma_z/2$ and the $\{u^{(j)}_{n\sigma,m}\}$ is the eigenvector characterizing 
the $m^\textrm{th}$ energy eigenstate in the $j=l+1/2$ subspace.

Note that this particular form of the Hamiltonian is similar to that of a Bogoliubov-de 
Gennes Hamiltonian of a harmonically-trapped spinless $p$-wave 
superconductor~\cite{Stone2008}.
Furthermore, the complex number $i$ in the complex factor $a$ can be absorbed in 
the even numbered eigenvector components so that numerically a tri-diagonal symmetric 
matrix can be diagonalized. 
In practice, a cut-off energy $E_c=\omega(l_\mathrm{max}+1) \gg E_F$ is introduced 
and finite matrices are diagonalized with basis states having less energy than the cut-off.
We checked that our numerical results for the the low-energy eigenvalues and eigenvectors 
are independent of the chosen cut-off. 

\begin{figure*}[!ht]
\includegraphics[scale=0.9]{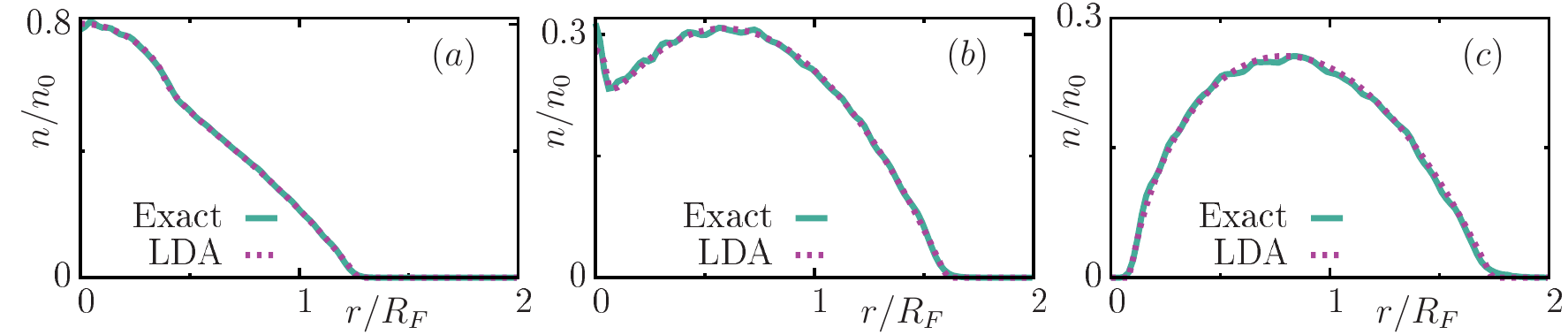}
\includegraphics[scale=0.9]{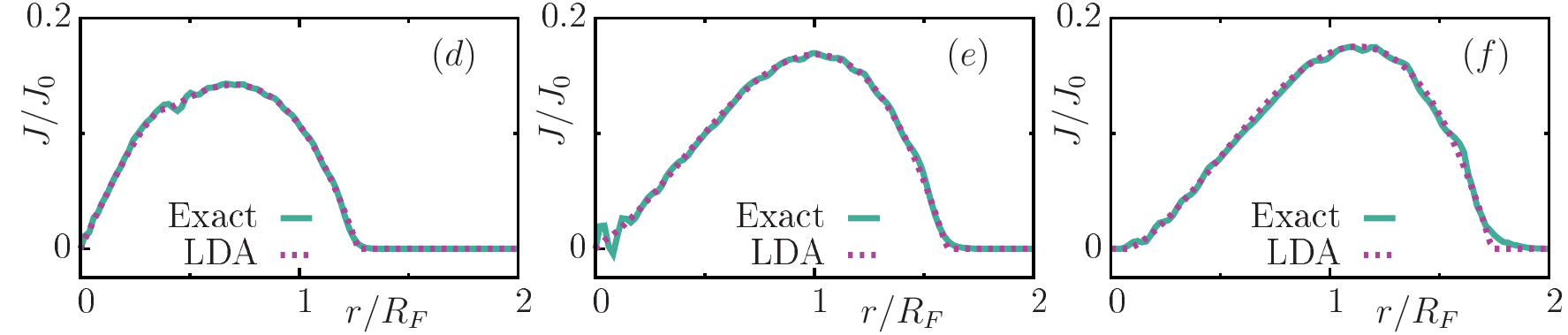}
\caption{(Color online) 
Exact solutions for the (a)-(c) number density, and 
(d)-(f) mass-current density are compared with those of semi-classical 
LDA approach, where $\Omega/\omega = 0.5$ in (a) and (d), 
$0.68$ in (b) and (e), and $0.7$ in (c) and (f). We also set $N=800$ 
and $\alpha=1.33 E_F/k_F$ in all figures. The overall results are in 
excellent agreement with each other up to minor deviations due to 
finite-size effects. 
\label{fig:LDA_vs_QM}
}
\end{figure*}

Since SOC couples $\uparrow$ and $\downarrow$ spins, the mass-current density 
should be identified from the continuity equation 
\begin{equation}
  \partial_{t}\rho(\br)+\mathbf{\nabla}\cdot\mathbf{J(\br)}=0,
  \label{eq:continuity}
\end{equation}
where 
$
\rho(\br)=M n(\br) 
$ 
is the mass density. The expectation value of the density and mass-current density 
operators are given as
\begin{eqnarray}
        n(r)   &=& 
       \sum_\sigma \langle \psi_{\sigma}^\dag(r) \psi_{\sigma}(r)\rangle 
       \nonumber \\
       &=& \sum_{n,n',l,l',\sigma} g^*_{n,l}(r) g_{n',l'}(r)\langle c_{n,l,\sigma}^\dag c_{n',l',\sigma}\rangle 
       \nonumber \\
       &=& \sum_{j} \sum_{m} 
       \left[ n^{(j)}_{\uparrow,m}(r) + n^{(j)}_{\downarrow,m}(r) \right] f\left[ E_{m}^{(j)} \right] , \\
       J_\theta(r)&=& \sum_\sigma \Im \langle \psi_{\sigma}^\dag(r) \frac{\partial}{r\partial\theta} \psi_{\sigma}(r)\rangle + 2M \alpha |\langle \psi_\uparrow^\dag(r)\psi_\downarrow(r)\rangle| \nonumber \\
	 &=&M \sum_{j} \sum_{m} J_{\theta, m}^{(j)}(r)  f\left[ E_{m}^{(j)}\right],
	\label{qmtotdens}
\end{eqnarray}
where 
$
f \left[ E_{m}^{(j)} \right] = \langle  C_{j,m}^\dag C_{j,m} \rangle
$ 
with $\langle \cdots \rangle$ the thermal average and $f(x)=1/(e^{\beta x}+1)$ is
the Fermi-Dirac distribution, and $\Im$ denotes imaginary part. Here, the density $ n^{(j)}_{\sigma,m}(r)$ 
and the angular component of the mass current density 
$J_{\theta, m}^{(j)}(r)$ for the $m^\textrm{th}$ energy eigenstate with total 
angular momentum $j$ are given by
\begin{align}
	n^{(j)}_{\uparrow,m}(r) &= 
	 \left\lvert \sum_{k=0}^{n_c/2} u^{(j)}_{l+2k\uparrow,m}g_{l+2k,l}(r,\theta) \right\rvert^2, \\
	n^{(j)}_{\downarrow,m}(r) &=
	 \left\lvert \sum_{k=0}^{n_c/2} u^{(j)}_{l+1+2k\downarrow,m}g_{l+2k,l+1}(r,\theta) \right\rvert^2, \\
	J_{\theta, m}^{(j)}(r) &= \frac{l}{r}n^{(j)}_{\uparrow,m}(r)+\frac{l+1}{r}n^{(j)}_{\downarrow,m}(r) \\
	 & + 2\alpha \left \lvert \left[\sum_{k=0}^{n_c/2} u^{(j)}_{l+2k\uparrow,m}g_{l+2k,l}(r,\theta)\right]^{*} \right. \\
	 & \left. \left[\sum_{k=0}^{n_c/2} u^{(j)}_{l+1+2k\downarrow,m}g_{l+2k,l+1}(r,\theta) \right] \right \rvert,
	\label{eq:qmdens1}
\end{align}
for $j>0$ ($l\ge 0$) and
\begin{align}
	n^{(j)}_{\uparrow,m}(r) &= 
	 \left\lvert \sum_{k=0}^{n_c/2} u^{(j)}_{-l+2k\uparrow,m}g_{l+2k,l}(r,\theta) \right\rvert^2, \\
	n^{(j)}_{\downarrow,m}(r) &=
	 \left\lvert \sum_{k=0}^{n_c/2} u^{(j)}_{-l-1+2k\downarrow,m}g_{l+2k,l+1}(r,\theta) \right\rvert^2, \\
	J_{\theta, m}^{(j)}(r)&=\frac{l}{r}n^{(j)}_{\uparrow,m}(r)+\frac{l+1}{r}n^{(j)}_{\downarrow,m}(r) \\
	& + 2\alpha \left \lvert\left[\sum_{k=0}^{n_c/2} u^{(j)}_{-l+2k\uparrow,m}g_{l+2k,l}(r,\theta)\right]^{*} \right. \\
	& \left. \left[ \sum_{k=0}^{n_c/2} u^{(j)}_{-l-1+2k\downarrow,m}g_{l+2k,l+1}(r,\theta)\right] \right \rvert,
	\label{eq:qmdens2}
\end{align}
for $j<0$ ($l<0$) and $n_c=(|l|_\mathrm{max}-|l|)/2$.
The density is independent of the angle $\theta$ and the radial component of the mass-current 
density is zero for the energy eigenstates.

In Fig.~\ref{fig:LDA_vs_QM}, we present the number and mass-current density 
profiles in the trap that are obtained from the exact calculation given above and the 
LDA approach described in the main text, showing an excellent agreement between 
the two. The total number of particles is $N=2\pi \int \textrm{d}r\, r n(r)=800$. The finite-size 
effects vanish in the thermodynamic limit when $N \to \infty$.

In comparing the exact quantum-mechanical calculations with LDA, we scale the radial 
distance by the Thomas-Fermi radius $R_F$ and the density by $n_0=k_F^2/(2\pi)$, 
which are defined via $E_F=k_F^2/(2M)=M\omega^2 R_F/2$. Here, the Fermi energy 
is defined for the non-rotating gas in the absence of an SOC as $E_F=\omega\sqrt{N}$ 
in LDA, so that $(k_Fa_0)^2 = 2 \sqrt{N}$ and $R_F/a_0=(4N)^{1/4}$. In this case, 
the exact quantum-mechanical solution gives $E_F=\omega(N_F+1)$ and 
$N=(N_F+1)(N_F+2)$ with the quantum number $N_F$. This is consistent with the 
LDA result $E_F=\omega\sqrt{N}$ for large number of particles $N$ in the trap such 
that $N_F\gg 1$. The mass-current density is in units of 
$J_0=Mn_0 \omega R_F=\sqrt{2}N^{3/4}/(\pi a_0^3)$.


\begin{thebibliography}{99}


\bibitem{Abo2001} J. R. Abo-Shaeer, C. Raman, J. M. Vogels, and W. Ketterle, 
Science \textbf{292}, 476 (2001).

\bibitem{Zwerlein2005} M.W. Zwierlein and J.R. Abo-Shaeer, A. Schirotzek, C.H.Schunck, W. Ketterle, 
Nature \textbf{435}, 1047 (2005).

\bibitem{Ho2000} T.L. Ho and C.V. Ciobanu, 
Phys. Rev. Lett. \textbf{85}, 4648 (2000).

\bibitem{Cooper2001} N. R. Cooper, N. K. Wilkin, and J. M. F. Gunn, 
Phys. Rev. Lett. \textbf{87}, 120405 (2001).

\bibitem{Fischer2003} U.R. Fischer and G. Baym, 
Phys. Rev. Lett. \text{90}, 140402 (2003).

\bibitem{Baranov2005} M. A. Baranov, K. Osterloh, and M. Lewenstein, 
Phys. Rev. Lett. \textbf{94}, 070404 (2005).

\bibitem{Hasan2010} M. Z. Hasan and C. L. Kane, 
Rev. Mod. Phys. \textbf{82}, 3045 (2010).

\bibitem{Qi2011} X.-L. Qi and S.-C. Zhang, 
Rev. Mod. Phys. \textbf{83}, 1057 (2011).

\bibitem{Sinova2004} J. Sinova, S. O. Valenzuela, J. Wunderlich, C. H. Back, and T. Jungwirth,
to appear in Rev. Mod. Phys. (2015).

\bibitem{Zutic2004} I. \v{Z}uti\'{c} and S. Das Sarma, 
Rev. Mod. Phys. \textbf{76}, 323 (2004).



\bibitem{Lin2011} Y.-J. Lin, Jim\'{e}nez-Garc\'{\i}a, and I. B. Spielman, 
Nature \textbf{471}, 83 (2011).

\bibitem{Zhang2012} J. Y. Zhang, S. C. Ji, Z. Chen, L. Zhang, Z. D. Du, Bo
Yan, G. S. Pan, B. Zhao, Y. J. Deng, H. Zhai, S. Chen, and J. W. Pan, 
Phys. Rev. Lett. \textbf{109}, 115301 (2012).

\bibitem{Wang2012}P. Wang, Z.-Q. Yu, Z. Fu, J. Miao, L. Huang, S. Chai, H. Zhai, and J. Zhang, 
Phys. Rev. Lett. \textbf{109}, 095301 (2012).

\bibitem{Cheuk2012} L.W. Cheuk, A.T. Sommer, Z. Hadzibabic, T. Yefsah, W.S. Bakr, and M.W. Zwierlein, 
Phys. Rev. Lett. \textbf{109},095302 (2012).

\bibitem{Qu2013} C. Qu, C. Hamner, M. Gong, C. Zhang, and P. Engels, 
Phys. Rev. A \textbf{88}, 021604(R) (2013).

\bibitem{Olson2014} A. J. Olson, S.-J. Wang, R. J. Niffenegger, C.-H. Li, C. H. Greene, and Y. P. Chen, 
Phys. Rev. A \textbf{90}, 013616 (2014).

\bibitem{Goldman2014} N. Goldman, G. Juzelinas, P. \"{O}hberg, and I. B. Spielman, 
Rep. Prog. Phys. \textbf{77}, 126401 (2014).

\bibitem{Jimenez-Garcia2015} K. Jim\'{e}nez-Garc\'{\i}a, L. LeBlanc, R.Williams, M. Beeler, C. Qu, M. Gong, C. Zhang, and I. Spielman, 
Phys. Rev. Lett. \textbf{114}, 125301 (2015).

\bibitem{Huang2015} L. Huang, Z. Meng, P. Wang, P. Peng, S.-L. Zhang, L. Chen, D. Li, Q. Zhou, and Jing Zhang,
arXiv:1506.02861 (2015).



\bibitem{Ruseckas2005} J. Ruseckas, G. Juzelinas, P. \"{O}hberg, and M. Fleischhauer, 
Phys. Rev. Lett. \textbf{95}, 010404 (2005).

\bibitem{Campbell2011} D. L. Campbell, G. Juzelinas, and I. B. Spielman, 
Phys. Rev. A \textbf{84}, 025602 (2011).

\bibitem{Xu2012} Z. F. Xu and L. You, 
Phys. Rev. A \textbf{85}, 043605 (2012).



\bibitem{Vyasanakere2011} J.P. Vyasanakere, S. Zhang, and V.B. Shenoy, 
Phys. Rev. B \textbf{84}, 014512 (2011).

\bibitem{Yu2011} Z.-Q. Yu and H. Zhai, 
Phys. Rev. Lett. \textbf{107}, 195305 (2011).

\bibitem{Iskin2011} M. Iskin and A. L. Suba{\c s}{\i}, 
Phys. Rev. Lett. \textbf{107}, 050402 (2011).

\bibitem{Gong2011} M. Gong, S. Tewari, and C. Zhang, 
Phys. Rev. Lett. \textbf{107}, 195303 (2011).

\bibitem{Yi2011} W. Yi and G.-C. Guo, 
Phys. Rev. A \textbf{84}, 031608(R) (2011).

\bibitem{Jiang2001} L. Jiang, X.-J. Liu, H. Hu, and H. Pu, 
Phys. Rev. A \textbf{84}, 063618 (2011).

\bibitem{Zhou2012} K. Zhou and Z. Zhang, 
Phys. Rev. Lett. \textbf{108}, 025301 (2012).

\bibitem{Liao2012} R. Liao, Y. Y.-Xiang, and W.-M. Liu, 
Phys. Rev. Lett. \textbf{108}, 080406 (2012).

\bibitem{Han2012} K. Seo, Li Han and C. A. R. S\'a de Melo, 
Phys. Rev. Lett. \textbf{109}, 105303 (2012). 



\bibitem{Takei2012} S. Takei, C.-H. Lin, B. M. Anderson, and V. Galitski, Phys. Rev.
A \textbf{85}, 023626 (2012).

\bibitem{He2012} L. He and X.-G. Huang, 
Phys. Rev. Lett. \textbf{108}, 145302 (2012).

\bibitem{Gong2012} M. Gong, G. Chen, S. Jia, and C. Zhang, 
Phys. Rev. Lett. \textbf{109}, 105302 (2012).

\bibitem{Ambrosetti2014} A. Ambrosetti, G. Lombardi, L. Salasnich, P. L. Silvestrelli, and F. Toigo, 
Phys. Rev. A \textbf{90}, 043614 (2014).

\bibitem{Yang2012} X. Yang and S. Wan, 
Phys. Rev. A \textbf{85}, 023633 (2012).

\bibitem{Zhang2013} W. Zhang and W. Yi, 
Nat. Commun. \textbf{4}, 2711 (2013).

\bibitem{Iskin2013} M. Iskin, 
Phys. Rev. A \textbf{88}, 013631 (2013).

\bibitem{Cao2014} Ye Cao, Shu-Hao Zou, Xia-Ji Liu, Su Yi, Gui-Lu Long, and Hui Hu,
Phys. Rev. Lett. \textbf{113}, 115302 (2014).



\bibitem{Martiyanov2010} K. Martiyanov, V. Makhalov, and A. Turlapov, 
Phys. Rev. Lett. \textbf{105}, 030404 (2010).

\bibitem{Dyke2011} P. Dyke, E. D. Kuhnle, S. Whitlock, H. Hu, M. Mark, S. Hoinka, M. Lingham, P. Hannaford, and C. J. Vale, 
Phys. Rev. Lett. \textbf{106}, 105304 (2011).

\bibitem{Frohlich2011} B. Fr\"{o}hlich, M. Feld, E. Vogt, M. Koschorreck, W. Zwerger, and M. K\"{o}hl, 
Phys. Rev. Lett. \textbf{106}, 105301 (2011).

\bibitem{Feld2011} M. Feld, B. Fr\"{o}hlich, E. Vogt, M. Koschorreck, and M. K\"{o}hl, 
Nature 480, 75 (2011).

\bibitem{Sommer2012} A. T. Sommer, L. W. Cheuk, M. J. H. Ku, W. S. Bakr, and M. W. Zwierlein, 
Phys. Rev. Lett. \textbf{108}, 045302 (2012).

\bibitem{Ries2015} M. G. Ries, A. N. Wenz, G. Zrn, L. Bayha, I. Boettcher, D. Kedar, P. A. Murthy, M. Neidig, T. Lompe, and S. Jochim, 
Phys. Rev. Lett. \textbf{114}, 230401 (2015).



\bibitem{Bausmerth2008} I. Bausmerth, A. Recati, and S. Stringari, 
Phys. Rev. Lett. \textbf{100}, 070401 (2008).

\bibitem{Bausmerth2008a} I. Bausmerth, A. Recati, and S. Stringari, 
Phys. Rev. A \textbf{78}, 063603 (2008).

\bibitem{Urban2008} M. Urban and P. Schuck, 
Phys. Rev. A \textbf{78}, 011601 (2008).

\bibitem{Iskin2009} M. Iskin and E. Tiesinga, 
Phys. Rev. A \textbf{79}, 053621 (2009).

\bibitem{Warringa2011} H. J. Warringa and A. Sedrakian, 
Phys. Rev. A \textbf{84}, 023609 (2011).

\bibitem{Warringa2012} H. J. Warringa, 
Phys. Rev. A \textbf{86}, 043615 (2012).

\bibitem{Radic2011} J. Radi\'c, T. A. Sedrakyan, I. B. Spielman, and V. Galitski,
Phys. Rev. A \textbf{84}, 063604 (2011).

\bibitem{Stone2008} M. Stone and I. Anduaga, 
Ann. Phys. (N. Y). \textbf{323}, 2 (2008).

\end{thebibliography}
\end{document}